\documentclass[twocolumn,aps]{revtex4}

\usepackage{graphicx}
\usepackage{dcolumn}
\usepackage{bm}
\usepackage[cmex10]{amsmath}

\begin{document}

\preprint{APS/123-QED}

\title{Spin-torque diode radio-frequency detector with voltage tuned resonance}

\author{Witold Skowro\'{n}ski}
 \email{skowron@agh.edu.pl}
\author{Marek Frankowski}
\author{Tomasz Stobiecki}
\affiliation{AGH University of Science and Technology, Department of Electronics, Al. Mickiewicza 30, 30-059 Krak\'{o}w, Poland}
\author{Jerzy Wrona}
\affiliation{AGH University of Science and Technology, Department of Electronics, Al. Mickiewicza 30, 30-059 Krak\'{o}w, Poland}
\affiliation{Singulus Technologies, Kahl am Main, 63796, Germany}

\author{Piotr Ogrodnik}
\affiliation{Faculty of Physics, Warsaw University of Technology, ul. Koszykowa 75, 00-662 Warsaw, Poland}
\affiliation{AGH University of Science and Technology, Department of Electronics, Al. Mickiewicza 30, 30-059 Krak\'{o}w, Poland}

\author{J\'{o}zef Barna\'s}
\affiliation{Faculty of Physics, Adam Mickiewicz University, ul. Umultowska 85, 61-614 Pozna\'n, and Institute of Molecular Physics, Polish Academy of Sciences,
Smoluchowskiego 17, 60-179 Pozna\'n, Poland}



\date{\today}

\begin{abstract}
We report on a voltage tunable radio-frequency (RF) detector based on a magnetic tunnel junction (MTJ). The spin-torque diode effect is used to excite and/or detect RF oscillations in the magnetic free layer of the MTJ. In order to reduce the overall in-plane magnetic anisotropy of the free layer, we take advantage of the perpendicular magnetic anisotropy at the interface between ferromagnetic and insulating layers. The applied bias voltage is shown to have a significant influence on the magnetic anisotropy, and thus on the resonance frequency of the device. This influence also depends on the voltage polarity. The obtained results are accounted for in terms of the interplay of spin-transfer-torque and voltage-controlled magnetic anisotropy effects.
\end{abstract}

\maketitle

Due to spin wave excitations, magnetic thin films can emit and/or absorb electromagnetic signals in a microwave frequency range determined by ferromagnetic resonance (FMR). In turn, radio-frequency (RF) signals can be produced in magnetic tunnel junctions (MTJs) {\it via} the spin-transfer-torque (STT) effect \cite{slonczewski_current-driven_1996, berger_emission_1996}, where a spin polarized current excites magnetization precession with a frequency that depends, among others, on the magnetic anisotropy. This precession gives rise to an electric signal that can be used in various devices, like spin wave generators \cite{demidov_direct_2010} or spin-torque oscillators \cite{kiselev_microwave_2003, skowronski_zero-field_2012}. The inverse effect, i.e.,  absorption of RF spin currents by a MTJ produces a detectable DC signal \cite{tulapurkar_spin-torque_2005}. This phenomenon, called the spin-torque diode effect, can be used as an RF-detector of very high sensitivity \cite{miwa_highly_2013}. Recently, it has been shown that similar behavior can be achieved with alternating-voltage-controlled magnetic anisotropy (VCMA) \cite{nozaki_electric-field-induced_2012, zhu_voltage-induced_2012}. In this paper we propose a voltage tunable spin-torque diode, that is capable of sensing RF signals of different frequencies. We achieved such a functionality by using a combination of the STT and VCMA effects. The former effect produces the DC voltage in response to AC current, while the latter one changes the resonance detection regime. In addition, using a specially designed MTJ, we are able to measure STT components in a voltage range of $\pm 1$ V, which to our knowledge is beyond the STT measurements published to date (up to $\pm 0.5$ V in Ref. \cite{wang_time-resolved_2011}) and reveals a non-linear behavior of the in-plane STT component vs. bias voltage \cite{wilczyski_free-electron_2008, jia_nonlinear_2011}.

\begin{figure}
\centering
\includegraphics[width=3.5in]{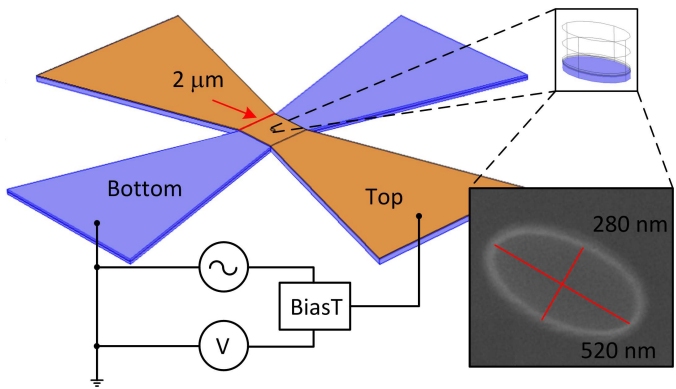}
\caption{The experimental  MTJ geometry. The long and short axes of the elliptical MTJ cross-section are 520 and 280 nm, respectively. The RF generator and voltmeter are connected using RF and DC connectors of the bias-T, respectively.}
\label{fig:Geometry}
\end{figure}

To achieve the above objective, we have investigated MTJs with the following multilayer structure: SiO$_2$ (substrate) / 5 Ta / 30 CuN / 3 Ta / 30 CuN / 3 Ta / 16 Pt$_{38}$Mn$_{62}$ / 2.1 Co$_{70}$Fe$_{30}$ / 0.9 Ru / 2.3 Co$_{40}$Fe$_{40}$B$_{20}$ / 1.6 MgO / 1-2 Co$_{40}$Fe$_{40}$B$_{20}$ / 10 Ta / 7 Ru  (thickness in nm). By varying thickness of the CoFeB free layer (FL), we observed a transition from in-plane to perpendicular anisotropy at a critical thickness of $t_c\approx 1.35$ nm \cite{skowronski_magnetic_2012}. After deposition, the MTJs were annealed at 250 $^{\circ}$C in an in-plane magnetic field of 0.4 T in order to set the exchange bias direction and to improve the crystallization of the ferromagnetic electrodes. For the transport measurements we used the MTJ with 1.6 nm thick FL, that reveals weak in-plane anisotropy. After patterning to the elliptical-shape pillars with the cross-section dimensions of 530 $\times$ 280 nm, using electron beam lithography, ion-beam etching and lift-off process, the TMR ratio of about 100 \% (measured at low bias voltage) and the resistance area (RA) product of 2 k$\Omega\mu$m$^2$ were achieved. All MTJ nanopillars were measured at room temperature in the in-plane and perpendicular magnetic field, and with a bias voltage applied with respect to the FL (top layer). High frequency measurements were performed using RF probes connected to the bias-T. The overlap between the bottom and top MTJ contact is 4 $\mu$m$^2$, which results in the MTJ capacitance of 4$\times$10$^{-15}$ F. The MTJ geometry and the nanopillar scanning electron microscope image are presented in Fig. \ref{fig:Geometry}.

Due to a high resistance of the MTJ of $R_P$ = 16 k$\Omega$ in the parallel magnetic state, the bandwidth $f_c$ = 1/(2$\pi$ $RC$) is limited to around 2.5 GHz. In addition, due to a resistance mismatch between our 50-Ohm measurement system, the RF reflection coefficient is $\Gamma$ = ($R$-50)/($R$+50) = 0.995. Thus, setting the power of the RF generator to 100 $\mu$W (-10 dBm), only 1\% (1-$\Gamma^2$) , i.e., 1 $\mu$W is applied to the MTJ, which results in I$_{RF}$ = 7.3 $\mu$A RF current and J$_{RF}$ = 5.9$\times$10$^7$ A/m$^2$ current density. We note that such current densities are sufficient to excite the magnetization dynamics by means of STT effect \cite{sankey_measurement_2007, skowronski_influence_2013}.

\begin{figure}
\centering
\includegraphics[width=\columnwidth]{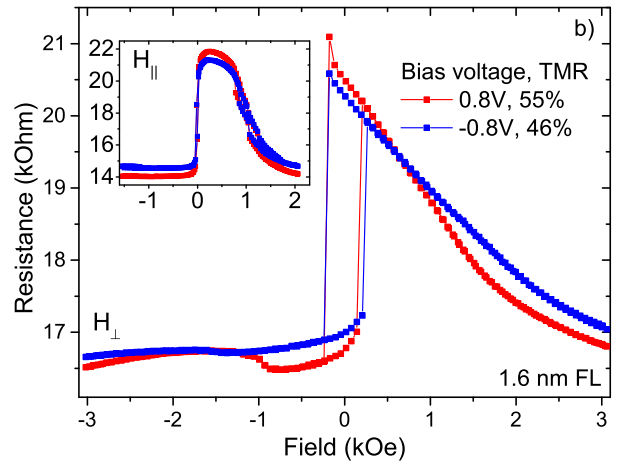}
\caption{Resistance vs. magnetic field loops measured for different bias voltage polarity and for the magnetic field applied perpendicular to the sample plane. Inset: the resistance-magnetic field loop for in-plane field.}
\label{fig:TMR}
\end{figure}

\begin{figure}
\centering
\includegraphics[width=\columnwidth]{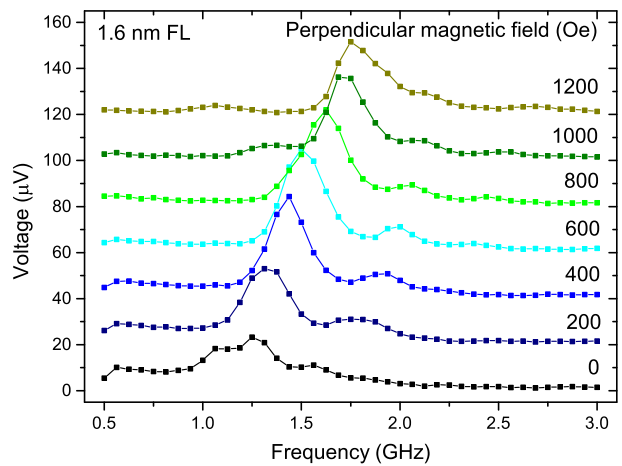}
\caption{FMR signal measured with a homodyne detection setup for a sample with 1.6 nm thick FL and for
different magnetic fields applied perpendicular to the sample plane. The successive curves are artificially offset for clarity.}
\label{fig:STFMR}
\end{figure}

Figure \ref{fig:TMR} presents resistance vs. in-plane and perpendicular magnetic field curves, measured for different bias voltage polarity. There is an evident influence of the bias voltage on the curves for perpendicular magnetic field.
Then, we performed spin-diode measurements on our MTJs. By applying perpendicular magnetic field, clear FMR spectra are observed,  Fig. \ref{fig:STFMR}, similar to those observed earlier in Ref. \cite{yakata_influence_2009}.

In order to model the dependence of the resonance frequency vs. perpendicular magnetic field we used a micromagnetic simulations using Object Oriented Micromagnetic Framework (OOMMF) \cite{oommf}. 
In the calculations we have included all ferromagnetic layers with Ru and MgO spacers. The whole five layer system was modeled using similar approach as in our previous work \cite{frankowski_micromagnetic_2013}. The cell size was set to 4 $\times$ 4 $\times$ 2 nm and the nanopillar cross-section was the same as in experiment. Interlayer exchange coupling between the FL and reference layer (RL) through the MgO barrier was set to zero, leaving only coupling due to a demagnetization field. Perpendicular anisotropy of the FL was equal to 870 kJ/m$^3$. The in-plane anisotropy was mainly determined by the shape anisotropy. However, in order to achieve a better agreement with the experiment, an additional magnetocrystalline in-plane anisotropy of 8 kJ/m$^3$ was assumed.

\begin{figure}
\centering
\includegraphics[width=\columnwidth]{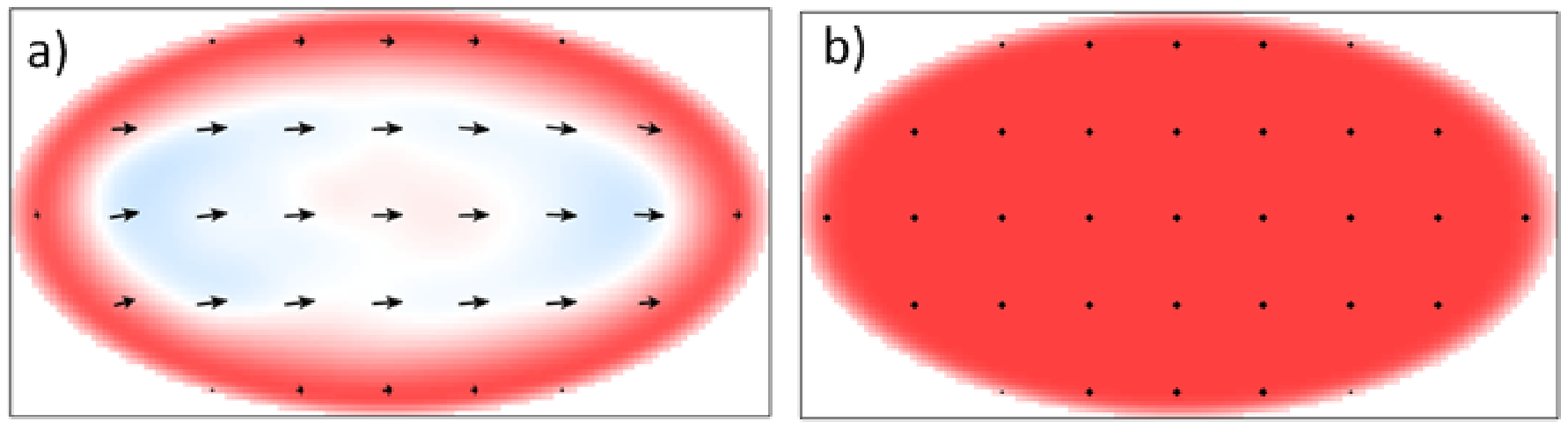}
\includegraphics[width=\columnwidth]{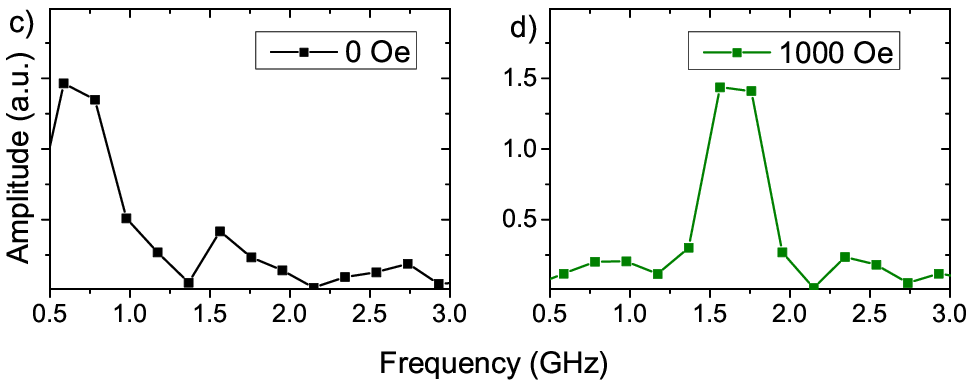}
\caption{Results of the micromagnetic simulations  of the FL magnetization distribution at (a) zero external magnetic field and (b) 1000 Oe perpendicular field. Note an inhomogeneous distribution (varying angle to plane) in the case (a). FMR curves simulated for the corresponding perpendicular magnetic fields are presented in (c) and (d). }
\label{fig:oommf_map}
\end{figure}

Due to an interplay between the in-plane shape anisotropy and strong perpendicular anisotropy at the CoFeB/MgO interface, the resulting magnetization of the FL at zero external magnetic field is inhomogeneous as presented in Fig. \ref{fig:oommf_map}a. For a perpendicular field of $H$ = 1000 Oe, the FL is evenly magnetized in a perpendicular direction, Fig. \ref{fig:oommf_map}b. The FMR spectra were simulated using similar approach as in Ref. \cite{skowronski_influence_2013}. Examples of OOMMF results obtained for perpendicular field of $H$ = 0 and 1000 Oe are depicted in Fig. \ref{fig:oommf_map}c-d. Note a nonuniform FMR spectrum for the inhomogenously magnetized FL (Fig. \ref{fig:oommf_map}c). The simulated resonant frequency for 1000 Oe field was close to the experimental one, \ref{fig:oommf_map}d. In order to achieve the frequency shift as measured with the bias voltage of 1 V, the perpendicular anisotropy constant had to be changed by 4 kJ/m$^3$.


\begin{figure}
\centering
\includegraphics[width=\columnwidth]{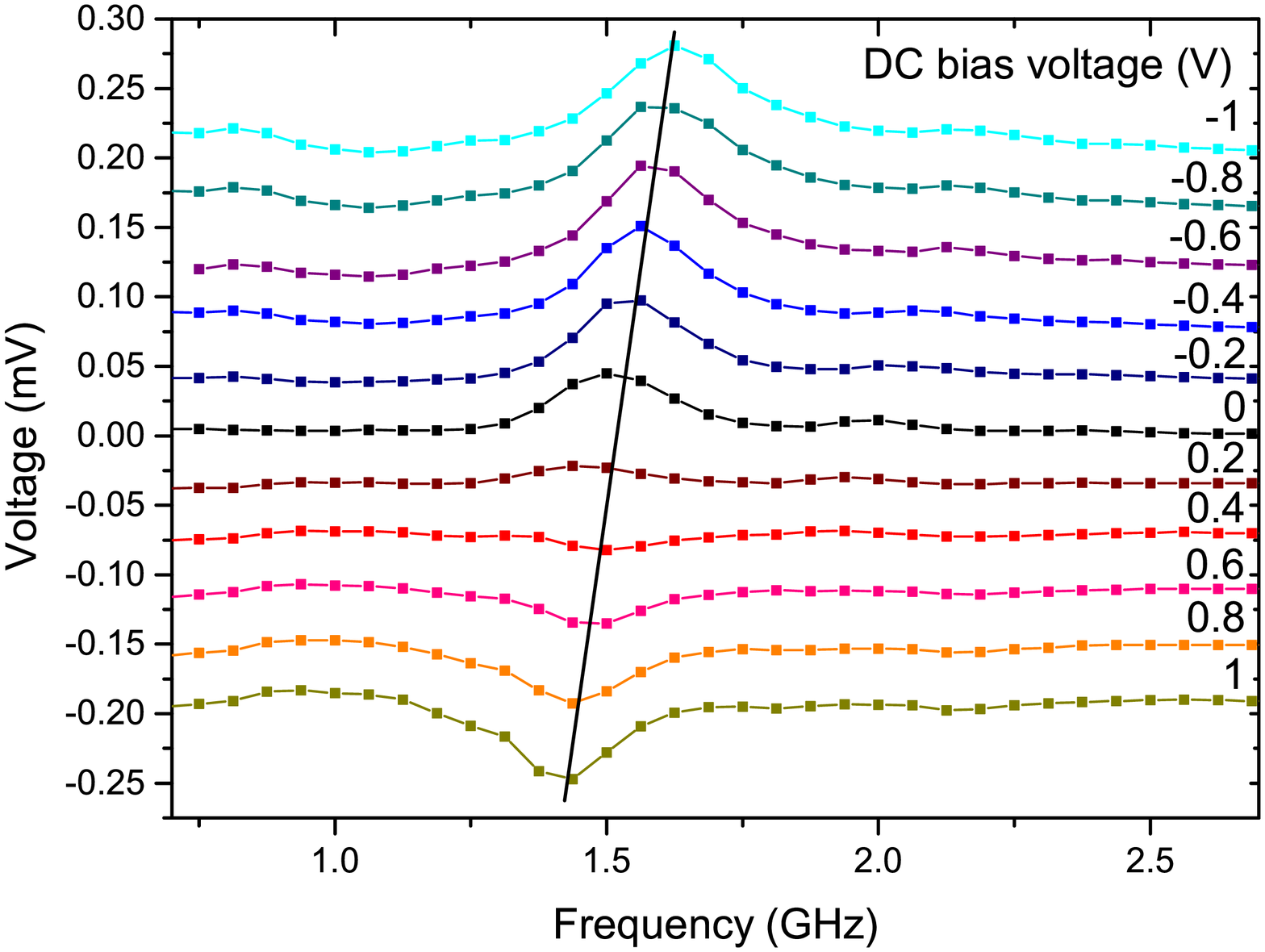}
\caption{FMR signal measured for perpendicular field of $H$ = 600 Oe in the bias voltage range from -1 $<$ V$_b$ $<$1 V. The black solid line is a guide for the eye, and represents a change in the frequency of -97 MHz/V.}
\label{fig:STD}
\end{figure}

From the TMR vs. magnetic field behavior we are able to calculate the angle $\theta$ between magnetizations of the FL and RL. We note, that for the field of $H$ = 600 Oe, the angle $\theta$ $\approx$ 90 $^\circ$, which results in maximum STT signal and minimum VCMA RF signal \cite{nozaki_electric-field-induced_2012}. Therefore, we attribute  the obtained mixing voltage signal to the STT-based phenomena. In addition, our signal exhibits pure Lorentzian shape, unlike dispersive signal in Ref. \cite{nozaki_electric-field-induced_2012}, which further supports the STT origin of the mixing voltage.
Next, we performed similar measurements for the magnetic field of $H$ = 600 Oe, but for different DC bias voltage in the range of -1 $<$ V $<$ 1 V. The obtained results are presented in Fig. \ref{fig:STD}. The ST-FMR signal inverses at around 0.3 V from positive to negative amplitude. Using the model presented in \cite{wang_bias_2009}, we have analyzed the obtained ST-FMR curves as a function of the bias voltage in order to derive the STT-components. 

\begin{figure}[!t]
\centering
\includegraphics[width=\columnwidth]{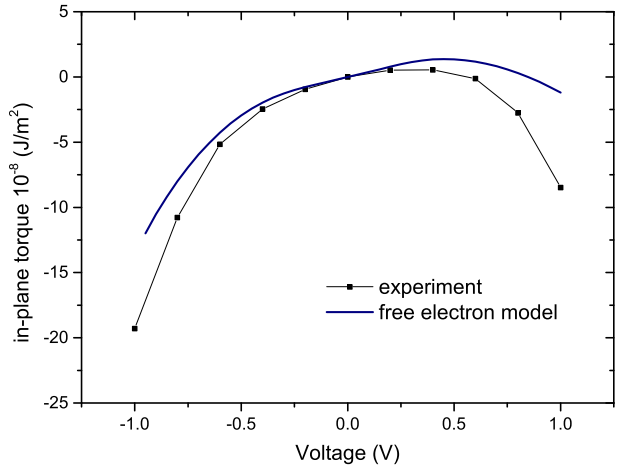}
\caption{
Comparison of the experimental in-plane torque with the theoretical one calculated within the  free-electron model.}
\label{fig:STT}
\end{figure}

Clearly, for all bias voltage values, the signal is of Lorentzian-shape. This results in the in-plane torque of two orders of magnitude larger than the out-of-plane torque \cite{skowronski_influence_2013}, contrary to the results measured typically in MTJs with thin (below 1 nm) MgO tunnel barriers. Based on this analysis, the in-plane torque in our MTJs exhibits parabolic-like shape in a larger voltage range, see Fig. \ref{fig:STT}.

The experimental results were compared with theoretical ones based on the free-electron model \cite{wilczyski_free-electron_2008}. As the first step we fitted the calculated current-voltage curves to the measured ones and we obtained a good qualitative agreement (not shown). Here, we used the band structure parameters corresponding to Co- and Fe-based electrodes: Fermi level (the width of the majority spins band) $E_{\mathrm{F}}$ = 1.1 eV and spin splitting energy $\Delta$\textit{E} = 1.3 eV. Such a structure corresponds to entirely polarized conduction band ($\Delta_1$) at zero bias voltage in accordance with the first principle calculations for Co(Fe)/MgO structures \cite{miller_dynamic_2007, yuasa_giant_2007}. In turn, the barrier was described by the potential $U$ = 1.3 eV (above the Fermi level) and the thickness of 1.6 nm. The height of the barrier was chosen arbitrary because of its great sensitivity to the interface structure \cite{yu_ab_2006}. However, this value seems to be reasonable and close to that measured experimentally for similar structures \cite{zaleski_study_2012}. Moreover, we introduced the effective electron mass $m^*_{\mathrm{MgO}} = 0.39m_0$ within the barrier, which is different from the electron effective mass in the electrodes, m$^*_{\mathrm{CoFe}}$ = m$_0$. An extension of the model introduced in Ref. \cite{wilczyski_free-electron_2008} requires an additional application of BenDaniel-Duke boundary conditions for wave functions and their first derivatives \cite{miller_dynamic_2007}. The presence of light carriers in the MgO layer reflects the fact that tunneling electrons (evanescent Bloch states) have different decay lengths within this kind of barrier, which depend on their symmetry \cite{fuchs_spin_2006}. As a result, we obtained a nonlinear torque dependence at higher bias voltage ( $\mid$V$\mid$ $>$ 0.4 V ), which agrees with the experimental curve. Contrary, for voltages $\mid$V$\mid$ $<$ 0.4 V, the experimental data as well as the theoretical ones exhibit linear dependence on the voltage. Similar behavior was observed for MTJs with thinner tunnel barriers, which however could not be probed at higher voltages. 


In order to exclude second and third order effects predicted in \cite{wang_bias_2009}, we repeated ST-FMR measurements for a different magnetic field resulting in theta close to 90 degrees, and we found the same behavior. Thus we could exclude non-linear second order FMR effects on our signal. The same experiment was performed on a few different MTJs of the same wafer, exhibiting similar results.

The STT-model, however, cannot account for the frequency peak shift with bias voltage. Such shift is measured in thin MgO MTJs as a result of perpendicular torque, which also changes the peak shape from a Lorentzian- to a dispersive-like curve \cite{sankey_measurement_2007}. Because we do not see any contribution form the perpendicular torque in a thick MgO MTJ, we can account the frequency shift with the bias voltage polarity to VCMA, see Fig. \ref{fig:STD}. Using the micromagnetic model presented above, we found that the change in the resonance frequency of -97MHz/V corresponds to the change in the perpendicular magnetic anisotropy  of 4 kJ/m$^3$/V. Such anisotropy changes are smaller than the values reported before \cite{shiota_quantitative_2011, endo_electric-field_2010, nozaki_voltage-induced_2010}. However, it was shown recently that both the direction and the magnitude of VCMA can depend strongly on the interface quality as well as on the adjacent layer \cite{shiota_opposite_2013}.


Taking into account the measured effects  and their interpretation based on the combination of STT and VCMA, we can design an RF detector that can be tuned in a certain frequency range. Changing the bias voltage polarity we can modify the peak-detection frequency from 1.6 to 1.4 GHz, i.e. in the range that could be interesting for wireless communication. The quality factor, though still low, can be improved using point-contact or vortex structures where the peak is typically of much smaller width \cite{skirdkov_influence_2012}. Furthermore, by increasing the device area, one can reduce the MTJ resistance, and thus, reduce the impedance mismatch and improve the sensitivity from the measured level of 11 to 1000 W/V.

In summary, we have investigated the spin-torque diode RF detector based on a MTJ with 1.6 nm MgO barrier.
The obtained FMR spectra have been accounted in terms of the STT effects. In turn, the difference in the FMR frequencies at different bias voltages has been attributed to VCMA. Based on our results we have also determined a non-linear in-plane torque vs. bias voltage in a wide voltage range.

\section*{Acknowledgement}
This work was supported by Polish National Science Center grant Harmonia-2012/04/M/ST7/00799. M.F. acknowledges Polish Ministry of Science and Higher Education Diamond Grant DI 2011001541. Numerical calculations were supported in part by PL-GRID infrastructure. J.B. acknowledges support by the Polish National Science Center as the Project No. DEC-2012/04/A/ST3/00372.

\bibliographystyle{nature}

\end{document}